\documentclass[default,iicol,sn-standardnature]{sn-jnl}% Default with double column layout

\jyear{2021}%

\theoremstyle{thmstyleone}%
\theoremstyle{thmstyletwo}%

\theoremstyle{thmstylethree}%

\begin{document}

\title[Neural Computing and Applications]{Graph Neural Netwrok with Interaction Pattern for Group Recommendation}


\author*[1]{\fnm{Bojie} \sur{Wang}}\email{BojieWang@bupt.edu.cn}

\author[1]{\fnm{Yuheng} \sur{Lu}}\email{yuheng.lu@bupt.edu.cn}

\affil[1]{\orgdiv{School of Computer Science (National Pilot Software Engineering School)}, \orgname{Beijing University of Posts and Telecommunications}, \orgaddress{\street{10 Xitucheng Road, Haidian District}, \city{Beijing}, \postcode{100876}, \state{Beijing}, \country{China}}}

% \affil[3]{\orgdiv{Department}, \orgname{Organization}, \orgaddress{\street{Street}, \city{City}, \postcode{610101}, \state{State}, \country{Country}}}


\abstract{With the development of social platforms, people are more and more inclined to combine into groups to participate in some activities, so group recommendation has gradually become a problem worthy of research. For group recommendation, an important issue is how to obtain the characteristic representation of the group and the item through personal interaction history, and obtain the group's preference for the item. For this problem, we proposed the model GIP4GR (\textbf{G}raph Neural Network with \textbf{I}nteraction \textbf{P}attern \textbf{For} \textbf{G}roup \textbf{R}ecommendation). Specifically, our model use the graph neural network framework with powerful representation capabilities to represent the interaction between group-user-items in the topological structure of the graph, and at the same time, analyze the interaction pattern of the graph to adjust the feature output of the graph neural network, the feature representations of groups, and items are obtained to calculate the group's preference for items. We conducted a lot of experiments on two real-world datasets to illustrate the superior performance of our model.}




\maketitle

\section{Introduction}\label{sec1}
    As the speed of information dissemination increases, more and more information begins to emerge in front of people. To solve the problem of the information overload and help people choose information of interest, recommendation systems have been widely deployed in online information systems, such as e-commerce platforms, social media sites, news portals, etc. The recommendation system can not only increase the traffic of service providers but also help customers find items of interest more easily. At present, most recommendation systems are for individual individuals to recommend, but as people's communication on social platforms becomes more convenient, people are more inclined to combine into groups to participate in activities. From this perspective, some studies have investigated recommending items to target user groups rather than individual users. This problem is called Group Recommendation \cite{ref1}. This form of users as a group is very common in online social media, users can organize into groups to participate in some online or offline activities.
    
\begin{figure}[ht]
        \centering
        \includegraphics[scale=0.38]{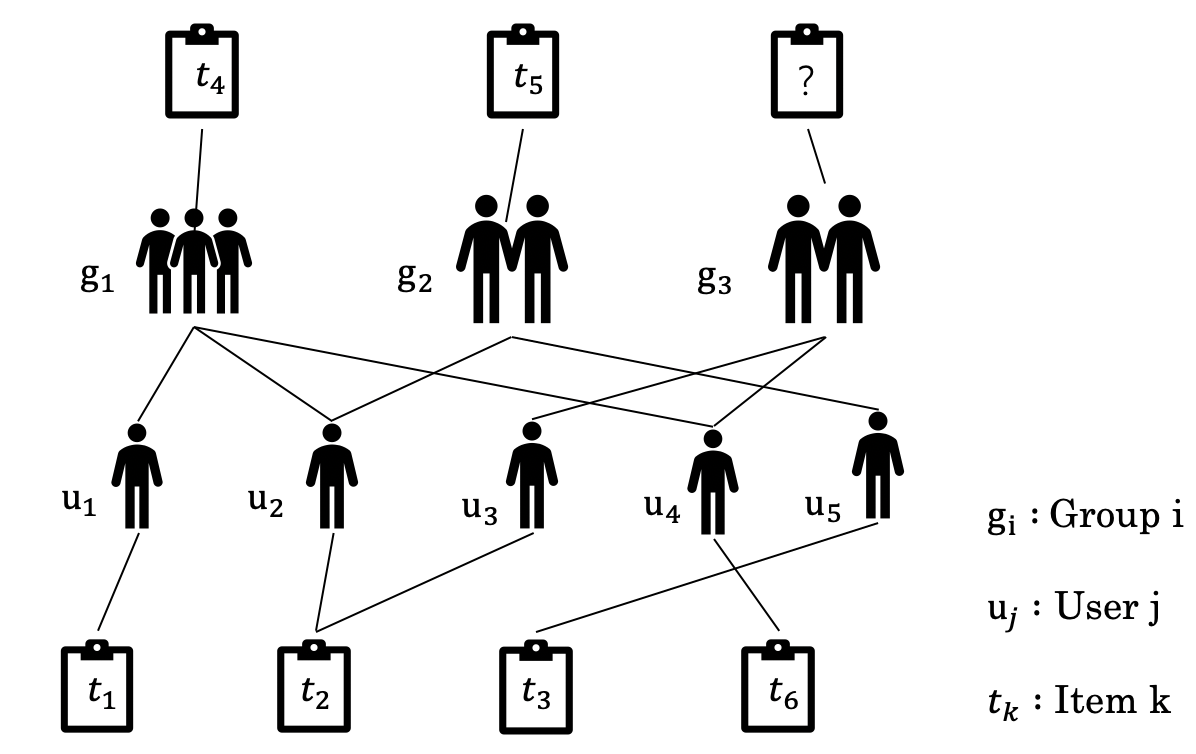}
        \caption{Group-User-Item interaction example}
        \label{Figure 1}
    \end{figure}

    \hyperref[Figure 1]{Figure 1} is a simple example of group interaction and individual interaction. In Figure 1, users u$_1$, u$_2$, u$_3$, u$_4$, and u$_5$ interacted with items t$_1$, t$_2$, t$_3$, t$_6$ respectively when they are individuals. When u$_1$, u$_2$, u$_4$ form g$_1$, they interact with items t$_4$. u$_2$, u$_5$ form g$_2$ interacted with item t$_5$. Our task is to predict the items to be interacted by group g$_3$ based on the interaction history of u$_3$ and u$_4$ when u$_3$ and u$_4$ form group g$_3$.
    
    Traditional recommendation methods are divided into model fusion and score fusion \cite{ref2}. The score-based fusion only simulates the group’s scores for items for each group member’s scores (specific methods can be divided into average scores, maximum scores, and minimum scores). This method only considers the group score arbitrarily. The relationship between the members and the final selection of items without further consideration of the influence of the members of the group on the decision. Model-based fusion is to train individual and item interaction models for each member of the group recently, and aggregate their preferences, so as to obtain a model as a group and then perform score prediction. This method can only be learned the abstract relationship between the group model and the individual model cannot measure the participation of different members in the final decision based on the interaction history of different members, and the model-based method needs to train the model for each group member, and compare the number of some users, for large datasets, the computational cost will be very large, so traditional recommendation methods cannot effectively perform recommendations.
    
    Recently, for the problem that previous work cannot effectively measure the participation of individuals in group decision-making, some scholars’ research on group recommendation focuses on how to automatically quantify the relative importance of individuals in group decision-making, that is, through the interaction of each member historical learning then determines the proportion of individuals in the final group decision to make trade-off decisions instead of using empirical strategies. For example, the attention mechanism is used to measure the importance of group members \cite{ref3,ref26}, the interaction history of each different group member is used to obtain different weights, and their weighted characteristics are used as the representation of the final group, thus making up for the traditional The shortcomings of recommendation strategies highlight the role played by individuals in group decision-making, and therefore exceed the performance of traditional group recommendation methods. At the same time, some scholars have proposed that the members of the group may have mutual influence \cite{ref4}, so a model is designed for each member to learn the interaction relationship between him and other members in the group, from his own and the other members of the group. The similarity (or influence) between others is used to update the individual’s characteristic representation, and finally, the representations of the members in the group are added to obtain the group’s representation. This method achieves quite good results, but the problem is similar to the traditional one based on the model fusion method is the same, that is, for a group with a large number of group members, it is expensive to train the model for each group member interacting with other group members. And these current methods are recommended based on the existing deep neural network methods. Although deep networks have powerful data fitting capabilities, they ignore the internal topological structure of the data and cannot learn the deep layers of the data more effectively. At the same time, group interaction data can be better realized by graph representation, so if a graph neural network is used, more specific features can be learned.
        
    In reality, many structures can be represented by graphs (such as social networks, protein structures, traffic networks, etc.), and the related technologies of graph neural networks have gradually matured \cite{ref5}. Therefore, graph neural networks have already had a wide range of applications. This also means that the abstract data structure can be expressed more vividly way, and the message passing method is adopted in the feature extraction, that is, a node can aggregate the information of the surrounding nodes, so it has a good effect on information aggregation. Given the advantages of graph neural network, this article uses the graph neural networks method to solve the problem of group recommendations. Some works have begun to use graph neural networks to study group recommendations \cite{ref6}, but only use the connectivity of the nodes of the graph to express the order of interaction, and do not go deep into the graph structure of group interaction. If a graph is used to represent the interaction between group users and items, the structure of the graph may be different depending on the dataset. For example, in some scenarios, when users form a group, they will choose to revisit the  previous place, that is, the items selected by the group will be a subset of the items the user has interacted with. At the same time, it is also possible that when a user forms a group, they will interact with an item that their group members have not interacted with. Therefore, in these two cases, the use of different layers of graph neural networks will lead to a large discrepancy in model performance.
    
    In general, the main contributions of this article are as follows:
    \begin{itemize}
        \item This article put forward the concept of interactive repetition rate(IRR), which distinguishes the interactive pattern of different items (such as movies, tourist attractions, restaurants, etc.), that is, how much users like to conduct group activities when they choose the ones they have interacted with.
        \item We propose a new group recommendation model GIP4GR based on a graph neural network and apply the proposed theory of behavior patterns to this model to form a universal model of group behavior.
        \item We conduct comprehensive experiments on two real-world datasets, and demonstrated the advantages of our model compared with the existing model, and verified some key designs of the model.
    \end{itemize}
    
    The structure of this paper is as follows: \hyperlink{sec2}{Section 2} summarizes the research status of group recommendation related work. \hyperlink{sec3}{Section 3} introduces the problem to be solved and the way the problem is expressed. \hyperlink{sec4}{Section 4} presents our solution to this problem. \hyperlink{sec5}{Section 5} introduces some details of the experiment and analyzes the results. Finally, \hyperlink{sec6}{Section 6} summarizes the paper and outlines the current shortcomings and future research directions.

\section{Related Works}\label{sec2}
    In this section we will introduce the problem and the related algorithms to solve the problem. Specifically, we will focus on current effort on group recommendation and the development of the graph neural network.
\subsection{Group Recommendation}\label{subsec2.1}
    Group recommendation requires that the personal preferences of all members in the group be properlyintegrated, so it requires a part of the process than a general personal recommendation.  Traditional grouprecommendation methods can be divided into memory-based and model-based methods.
    
    The first is the memory-based method. O'Connor et al. \cite{ref1} have proposed some traditional methods to aggregate recommendation results based on scores, such as maximum satisfaction (the highest score of a member in a group is selected as the group score for an item, in order to maximize group satisfaction), minimum pain (for an item, select the lowest score of a member in the group as the group score, thereby pleasing all members in the group), average (for an item, select the average score of the members in the group as the group's final score, thus weighing the maximum satisfaction and minimum pain) and other methods \cite{ref7}, however, these strategys is too empirical and too intuitive, ignoring the relationship within the group (for example, different members have different decision-making processes contribution). At the same time, each group within each dataset may have different aggregation methods, that is, some groups tend to be the most satisfied, some tend to be the least painful, and it is impossible to generalize to use a standard.
    
    The second is model-based methods \cite{ref9,ref10} mostly using pre-processing methods. Modeling the interaction between individuals and items in the group, and then fusing these user models to obtain a model representing the group, and then calculating the score of the interaction between the group and the item, so only relatively shallow features are learned. However, groups in real life change frequently, and the members of the group may also partially overlap, which are only formed for very few activities. Moreover, it is not possible to refine the participation of each member in the group.
    
    At the same time, the probability model is also applied to solve the group recommendation. The PIT model \cite{ref10} proposes to represent the entire group as the user with the greatest influence in the group, but ignores the fact that the model will be effective only when the user with the greatest influence is an expert in this area, otherwise, the group internal attention will also shift in this respect, that is, the influential ones will listen to the experienced ones.
    
    Recently, in order to measure the percentage of participation of members in the group based on the interactions that the group has previously participated in, a deep representation model based on learning has been proposed \cite{ref4,ref3,ref13}, all of which use the attention mechanism \cite{ref12}, to measure the weight of different members of the group. It turns out that they perform better than models based on score aggregation or model aggregation.

\subsection{Application of graph neural network in recommendation system}\label{subsec2.2}
    Algorithms for extracting graph information are also emerging in endlessly. For example, GCN \cite{ref14} weights and aggregates the information of surrounding nodes for each node in the graph according to the in-degree and out-degree of the node; GraphSAGE \cite{ref15} is to sample the surrounding nodes of a node and then aggregate and provide a variety of aggregation methods, such as Max-Pooling, LSTM, Mean-Pooling; GAT \cite{ref16} is to calculate the information around each node, get the weight of each node around, and then obtains the weighted central node information. These methods can be effectively applied to the recommendation system. Most of the recommendation system models introduced before are based on general deep neural networks, etc., and do not take advantage of the powerful representation capabilities of graph neural networks. The relationship between groups and users can be represented by graphs for more significant learning, to the characteristics of the group. The application of graph neural network in recommendation system includes: Pin-Sage \cite{ref17}is a kind of spread representation on the project graph through the graph convolutional layer. NGCF \cite{ref18} models the bipartite graph of user items, so as to learn to aggregate the interactive information of user items. RippleNet \cite{ref19} uses a knowledge graph to aggregate interactive items and richer personal information and item information. At the same time, GLS-GRL \cite{ref6} also uses graph neural networks for the problem of sequential group recommendation, that is, graph neural networks are used to represent long-term and short-term user-item interactions, so as to learn how group interests change over time. SR-GNN \cite{ref5} uses a graph neural network to learn the sequence relationship of items interacted in a session to generate embedding representation and obtains the score of each item through attention mechanism and linear transformation.
    
    It can be seen that the application of graph neural networks in recommendation system have been widely used. However, for some datasets with different behavior patterns, the graph neural network with a fixed number of layers is easy to show over-smoothness \cite{ref20} or under-fitting phenomenon resulting in poor recommendation performance, that is, GNNs that need to adjust the model artificially for different datasets. To solve this problem, we dynamically learn the group interaction mode to obtain the interaction repetition rate(IRR) of the group to give different importance to the output of different layers as the final feature representation.

\section{Annoation And Problem Formulation}\label{sec3}
    In this section we will formulate the problem we will sovle and the symbol we will use in the following sections.
    
    \begin{table}[h]
    \begin{center}
    \begin{minipage}{200pt}
    \caption{Notation}\label{tab1}%
    
    \begin{tabular}{@{}l l@{}}
    \toprule
    Notation    &     Description\\
    \midrule
    G, U, T       &     The set of group, user, item\\
    g$_i$, u$_j$, t$_k$ &     The i-th group, j-th user, k-th item\\
    g$^{(x)}_i$, u$^{(x)}_j$&    The x-th item of i-th group and j-th user \\ &interacted respectatively\\
    \textbf{e}$^{(n)}_z$          &    Output embedding of node z(user, group \\&or item) from n layer of model\\
    $\mathcal{G}, \mathcal{V}, \mathcal{E}$ & input set of graph, nodes, edges\\
    \textbf{x}$_z$ & Input embedding of node z(user, group \\& or item)\\
    $type()$         & Type of the entity, collection operation\\
    \botrule
    \end{tabular}
    \end{minipage}
    \end{center}
    \end{table}
    Suppose we have N users \textit{U}=\{u$_1$, u$_2$,..., u$_n$ \}, S groups G=\{g$_1$, g$_2$,..., g$_s$\}, and M items \textit{T}=\{t$_1$, t$_2$,. .., t$_m$\}. The i-th group g$_i$ $\in$ G contains members \{u$_{1}$, u$_{2}$...\}. There are three interaction relationships in the dataset, namely group-user, group-item, and user-item. The entire dataset can be seen as an interact graph, where the items, users, groups, can be seen as the nodes of the graph and their interaction-ship can be seen as the edge of the graph.
    
    Problem: For a specific group, generate a corresponding Top@N recommended item list.
    
    Input: user set \textit{U}, group set \textit{G}, item set \textit{T}, users, items, groups'interact graph.
    
    Output: A list of items generated for group g$_i$, \textit{T'}=\{t$_1$, t$_2$,...\}.

\section{Methology}\label{sec4}
\begin{figure*}[ht]
        \centering
        \includegraphics[scale=0.4]{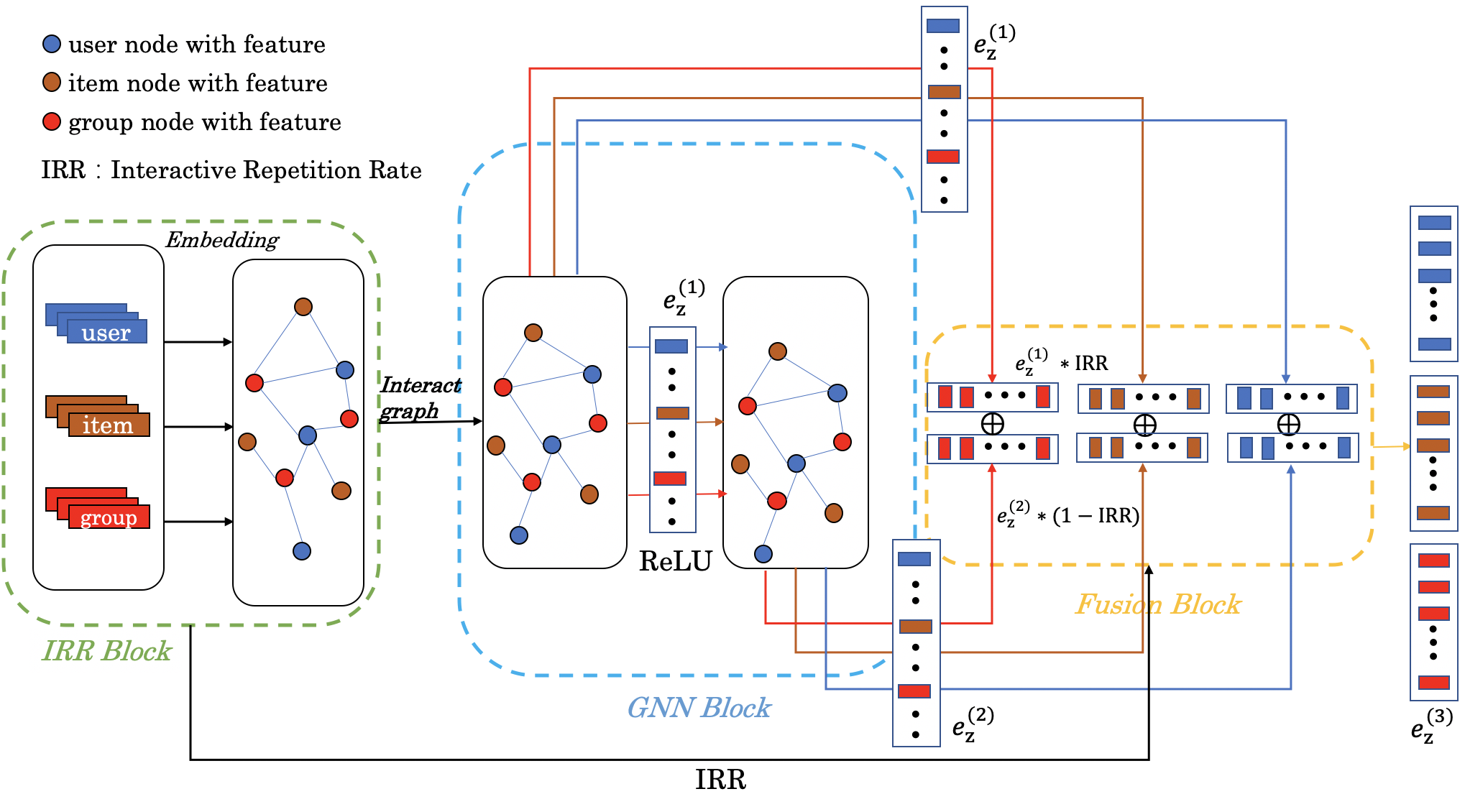}
        \caption{Framework of GIP4GR.}
        \label{Figure 2}
    \end{figure*}

    In this part we will introduce component of the model and training details.
    The overall structure of the model is shown in \hyperref[Figure 2]{Figure 2}. In general, our proposed model GIP4GR includes three parts. 1) Calculating the interactive repetition rate (\hyperref[subsec4.1]{section 4.1}). The green dashed box (IRR Block) in \hyperref[Figure 2]{Figure 2} is mainly used to calculate the interactive repetition rate according to the behavior characteristics of different datasets. 2) Learning interactive information (\hyperref[subsec4.2]{section 4.2}) is the blue dashed box (GNN Block) in \hyperref[Figure 2]{Figure 2}, which is the part of applying graph neural network to learn the in-depth information of the topological structure of the dataset. 3) Get the final representation (\hyperref[subsec4.3]{section 4.3}) in \hyperref[Figure 2]{Figure 2} is a yellow dashed box (Fusion Block), according to different interaction repetition rates and interaction information to get the final group, user, and item embedding. Training details are illustrated in \hyperref[subsec4.4]{section 4.4}.

    \subsection{Calculate the Repetition Rate of Group Interaction}\label{subsec4.1}
        This part corresponds to the green dashed frame block in \hyperref[Figure 2]{Figure 2}, mainly for calculating IRR indicators and applying them to subsequent calculation tasks.
        
        \begin{figure}[ht]
        \centering
        \includegraphics[scale=0.36]{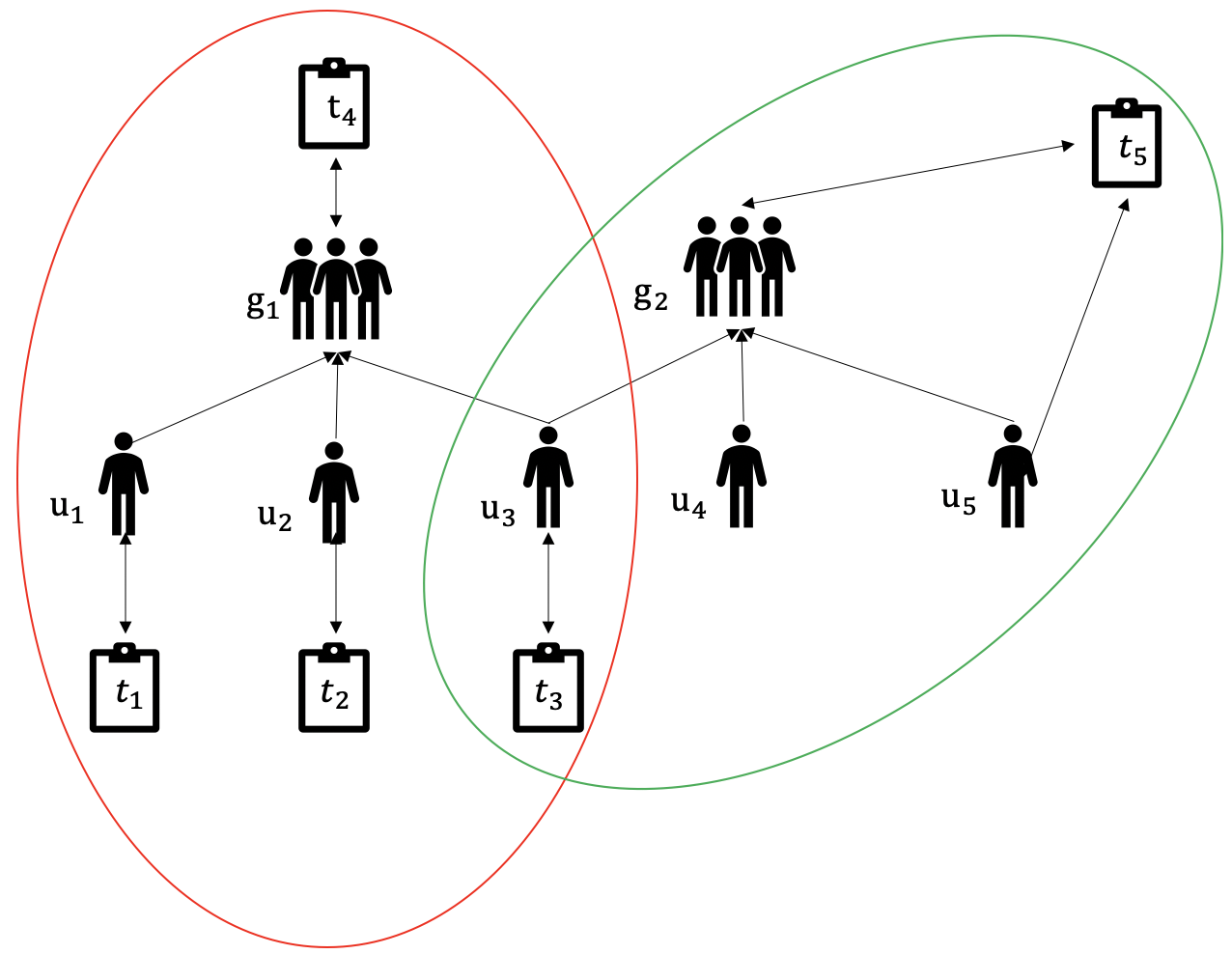}
        \caption{As a special case of group interaction, it is explained here that when users form a group, they can interact items that have been interacted before or items that have not been interacted. For example, group g$_1$ interacts item t$_4$ that has not been interacted by members of the group, and group g$_2$ interacts item t$_5$ that has been interacted by u$_5$.}
        \label{Figure 3}
        \end{figure}
        
        In daily life, the scenarios in which we form groups and then choose items to interact can be divided into the following two types:
        
        As shown by the red circle in \hyperref[Figure 3]{Figure 3}, in certain scenes (such as traveling) when u$_1$, u$_2$, and u$_3$ are organized into group g$_1$, they choose t$_4$ which they have not experienced before. That is, when people organize into groups, they may get tired of what they have experienced before or organize into groups to participate in projects suitable for group activities. \hyperref[Figure 1]{Figure 1} is also the usual situation.
        
        The situation described by the green circle in \hyperref[Figure 3]{Figure 3} is just the opposite. When u$_3$, u$_4$, and u$_5$ are organized into a group g$_2$, they choose the t$_5$ item that u$_5$ has experienced before, and then the item t$_2$ that u$_3$ has experienced before does not work. For example, in a watching movie scenario, users like spend time and energy to form a team to 
        watch the movie they have watched. This leads to different interaction 
        
        patterns for different datasets so that the fixed number of GNN layers in the model cannot reach a certain value.
        In order to measure the extent to which the used dataset belongs to which of these two situations, we propose an indicator of \textbf{Interactive Repetition Rate (IRR)}:
        
        \begin{equation}
            \begin{aligned}
            IRR=\frac{1}{S}\sum_{i=0}^{S}\frac{type(t^{(1)}_i,t^{(2)}_i...t^{(n)}_i)\cap type(t^{(1)}_j,t^{(2)}_j...t^{(m)}_j)}{type(t^{(1)}_i,t^{(2)}_i...t^{(n)}_i)} \\, j\in g_i 
            \end{aligned}
        \end{equation}
        
        IRR is used to adjust the weight of the output of different layers of the network in the final output, which solves the problem of the group interaction type and the number of layers of the graph neural network.
    \subsection{Feature Representation Learning}\label{subsec4.2}  
        This part corresponds to the blue dotted box block in \hyperref[Figure 2]{Figure 2}, which uses the current mature GraphSAGE and GAT to aggregate and learn the characteristics of user nodes, project nodes, and group nodes.
        
        \begin{figure*}[ht]
        \centering
        \includegraphics[width=\textwidth]{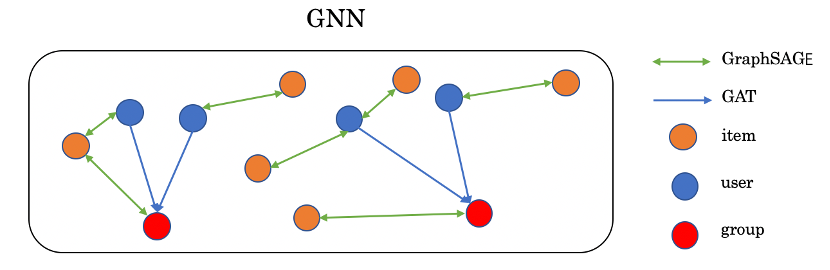}
        \caption{IRR block structure.}
        \label{Figure 4}
        \end{figure*}
        
        The partial structure of the GNN is shown in \hyperref[Figure 4]{Figure 4}. We use the sampling and aggregation framework GraphSAGE to aggregate the group-item and user-item (bidirectional edges) interactions. The formula is as follows:
        
        \begin{equation}
        \boldsymbol{{e'_u}^{(l)}}=AGGREGATE_{aggtype}\{\boldsymbol{e_t^{(l)}},\forall t\in N_u,N_u\in T\}\nonumber
        \end{equation}
        \begin{equation}
        \boldsymbol{{e_u}^{(l+1)}}=\sigma(W^{(l+1)}·[\boldsymbol{{e_u}^{(l)}},\boldsymbol{{e'_u}^{(l)}}])
        \end{equation}
        
        Where \textit{aggtype} in this paper that is the Max-Pooling method. $\sigma$ means the sigmoid activation, $W^{(l)}$ is the learning parameter matrix of the l layer, $N_u$ represents the one-hop adjoining node of u, the feature of iteration is that the item t nodes around the user node u are aggregated in the formula, and the aggregation of item nodes and group nodes is also the same.
        
        In the group-user interaction, we use a one-way edge (from the user to the group), and use the GAT convolution method on this type of edge, aiming to obtain different attention weights for users according to different user characteristics. The formula is as follows:

        \begin{equation}
        \begin{aligned}
        \alpha_{gu}=\frac{exp(actf(a^T[W^{(l)}\boldsymbol{e_g^{(l)}}||W^{(l)}\boldsymbol{e_u^{(l)}]))}}
        {\sum_{k\in N_g,N_g\in U}exp(actf(a^T[W^{(l)}\boldsymbol{e_g^{(l)}}||W^{(l)}\boldsymbol{e_k^{(l)}}])))}
        \end{aligned}
        \end{equation}
        
        Among them, actf means the activation function, here we use the LeakyReLU in GAT, $\alpha_{gu}$ represents the proportion of member u in group g. We only aggregate the user’s characteristics into the group, that is, the user does not aggregate group information. This is to allow the group to use the user’s historical interaction information. Learn the information related to the final decision of the group. Here we use only one attention head in GAT (experiments have shown that one head can achieve good results and also reduces computational overhead).
        
        It is worth noting that these five edge convolution methods are all performed at the same time, and the nodes that have been repeatedly convolved are added. (For item t$_5$ in  \hyperref[Figure 3]{Figure 3}, the information of g$_2$ and u$_5$ will be gathered at the same time, and g$_2$ the information of the person is added), that is, all nodes are only equivalent to applying a layer of GNN.
    \subsection{Multilayer GNN}\label{subsec4.3}
        This part corresponds to the Fusion Block in the yellow dashed frame in \hyperref[Figure 2]{Figure 2} and uses the interactive repetition rate to obtain the weighted sum of the network output of different levels as the final feature representation.
        
        As shown by the red and green circles in \hyperref[Figure 3]{Figure 3}, the datasets of different group interaction repetition rates have different graph diameters, so it is very important to choose the number of layers of the graph network.
        
        If the number of selected layers is too small, the embedded representation ability will be insufficient, that is, the group representation cannot fully obtain the necessary information of its surrounding nodes, e.g.: Assume that the items selected by the group are not interacted by the group members before. If only one layer of the network is used, the group can only aggregate the information of its members and the information of the items selected by the group, but cannot learn the information of the historical interaction items of the group members, which means that it is impossible to learn from the interaction of a single group member to the aggregation.
        
        If the number of layers selected is too large, it may cause an over-smoothing phenomenon \cite{ref20}, that is, the information of the group and the item is similar, e.g.: Assuming that the item selected by the group is previously interacted by a member of the group, then if using  two layer GNN will first cause the group to interact the item and the group members who have interacted with the item (as shown in g$_2$, u$_5$, and t$_5$ in \hyperref[Figure 3]{Figure 3}), the group g$_2$ will be aggregated twice u$_5$'s information, twice t$_5$'s information, u$_5$ and t$_5$ are also similar), the information has been spread many times, resulting in the three expressions being very similar, and secondly, some team members (not enough to contribute to the final decision) may have historical interactions information is learned, resulting in information redundancy, and it is difficult to assign different weights to group members.
        
        Therefore, we propose to use the previously calculated group interaction repetition rate as a trade-off. We propose to use the two layer GNN model mentioned before for feature learning (from \hyperref[Figure 3]{Figure 3}, it can be seen that if the group needs to learn the final feature, the maximum diameter is item (group selection)-group-member-item (Member selection), that is, it takes two hops to get the group member’s interest information and the item information to be aggregated into the group, so the maximum number of layers is set to two), and we use the interactive repetition rate to neutralize the first and second layers The output of is used as the final representation. That is, if the interactive repetition rate is relatively large, it means that the information output by the first layer of the graph network is more important (only the information of the surrounding nodes of the first hop of the node is aggregated), so we use \textit{IRR} to represent the interactive repetition rate, then there is:
        
        \begin{equation}
            \boldsymbol{e^{(3)}_z}=IRR*\boldsymbol{e^{(1)}_z}+(1-IRR)*\boldsymbol{e^{(2)}_z}, z\in (U,G,T)
        \end{equation}
        
        In this way, the final embedding can dynamically adjust the proportion of the output of the first and second layer network in the final output according to the group interaction.
        
        This idea mainly comes from the classic computer vision algorithm: residual connection \cite{ref21}, which is used to solve the problem of gradient disappearance and gradient explosion on the back layer network. It can also refer to JKnet \cite{ref22}, which is a method specifically used to solve the problem of excessive smoothing caused by too many layers in the graph network. JKnet stitches the output of all layers at the end, Max-Pooling and LSTM operations to obtain The final representation, and we are based on the way the group interacts, and it has to be weighted for specific purposes.The specific process of model is \hyperref[alg1]{Algorithm 1}:
        
        \begin{algorithm}[ht]\label{alg1}
        % \hline
            \caption{\textbf {The Method of GIP4GR}}  
            % \SetAlgoLined
            {Input:$\mathcal{G}(\mathcal{V},\mathcal{E})$, Nodes: $\forall\mathcal{V}_{user},\mathcal{V}_{item},\mathcal{V}_{group}\in\mathcal{V}$, Nodes'feature: $\textbf{x}_v,\forall v \in \mathcal{V}$}\\
            % \KwResult{Vector representations for each of users,items,groups}\\
            % \vbox{}
            Initialization;
            
            \textbf{Step1:} Caculate the IRR\\
            % IRR=$\frac{1}{S}\sum_{i=1}^{S}\frac{type(t_{i1},t_{i2}...t_{in})\cap type(t_{j1},t_{j2}...t_{jk})}{type(t_{i1},t_{i2}...t_{in})}, j\in g_i$\;\\
            IRR=$\frac{1}{S}\sum_{i=0}^{S}\frac{type(t^{(1)}_i,t^{(2)}_i...t^{(n)}_i)\cap type(t^{(1)}_j,t^{(2)}_j...t^{(m)}_j)}{type(t^{(1)}_i,t^{(2)}_i...t^{(n)}_i)}, j\in g_i$ \;\\
             \vbox{} 
            \textbf{Step2:} Get The representation of nodes from all layers of the GNNs\\
            $\boldsymbol{e^{(1)}_z}=GNN_1(\mathcal{G},\textbf{x}_v);$\\
            $\boldsymbol{e^{(2)}_z}=GNN_2(\mathcal{G},\boldsymbol{e^{(1)}_z}),(z\in \mathcal{V});$
            
            \textbf{Step3:} Get the final nodes representation\\
            $\boldsymbol{e^{(3)}_z}=IRR*\boldsymbol{e^{(1)}_z}+(1-IRR)*\boldsymbol{e^{(2)}_z}, z\in (U,G,T)$\\
        % \hline
        \end{algorithm}
    \subsection{Model Training and Optimization}\label{subsec4.4}
        \subsubsection{GIP4GR Prediction Method}
                We use the classic problem type in GNN link prediction \cite{ref23}, which is to judge whether there will be an edge between the target group and the target item
                 {}
                \begin{equation}
                    \check{y}_{g, v}=\phi(\boldsymbol{e^{(3)}_g}, \boldsymbol{e^{(3)}_v})
                \end{equation}
                 {Among them,$\check{y}_{g, v}$represents the probability(or the size of the score) that g and v have edges, $\boldsymbol{e^{(3)}_g,e^{(3)}_v}$ represents the final representation of g and v respectively, $\phi$ represents the prediction function, it has many forms like dot product, MLP, cosine similarity etc. In our experiment, our prediction function uses the dot product method (the more similar the same subspace, the larger the score), so the predicted result is the score between the items recommended for a group.}
            \subsubsection{Model Optimization}
                Regarding group recommendations, we provided display feedback based on negative samples. Based on this, the score of the observed interaction should be higher than the corresponding score of the unobserved for optimization. Our loss function is as follows:
                
                \begin{equation}
                    arg \mathop{min}_{\ominus}\sum_{(i,k,k')\in {\textit{Train}}}\{1-\sigma(\check{y}_{i,k}-\check{y}_{i,k'})\}
                \end{equation}
                
                 {\textmd{Where \textit{Train} represents the training set, that is, the group item interaction graph and the (i, k, k') triplet indicates that the group i has interacted with k items but has not interacted with k' items (we take the items that have not interacted with the group as negative samples ), where \textbf{$\sigma$} represents the sigmoid function, $\ominus$ means the parameters of model. The main purpose of this is to widen the gap between the scores of the positive samples and the negative samples so that the features between the positive samples and the negative samples are prominent.}}

\section{Experiment}\label{sec5}
    In this section, we conducted a lot of comparative experiments with the current benchmark model on two real-world datasets and answer the following research questions:
    \begin{itemize}
        \item \textbf{RQ 1:}\label{RQ 1} Compare our proposed model with the existing models, whether the recommendation performance is better?
        \item \textbf{RQ 2:}\label{RQ 2} Can the experiment prove that including the group interaction repetition rate has an impact on the performance of the model recommendation, that is, can it solve the under-fitting and over-smooth phenomenon when applying graph networks?
    \end{itemize}
    
    \subsection{Experiment Setting}
        \subsubsection{Datasets}
            Due to the previously mentioned dataset \cite{ref6,ref7} may not meet the behavioral characteristics shown in Section 3.2, we reused the two real-world datasets used in \cite{ref3,ref24} datasets CAMRa2011 and MaFengWo. The details of the datasets can be seen in \hyperref[Table 2]{Table 2}. It can be seen that the interactive repetition rate of CAMRa2011 and MaFengWo is exactly at the opposite extremes.

            \begin{table}[h]
            \begin{center}
            \begin{minipage}{300pt}
             \caption{\textbf{Illustration of Datasets}}
                \begin{tabular}{@{}l c c@{}}
                    \hline
                    Data                              &  CAMRa2011  &   MaFengWo\\
                    \hline
                    Number of Users                   &  602        &   5275\\
                    \hline
                    Number of Groups                  &  290        &   995\\
                    \hline
                    Number of Items                   &  7710       &   1513\\
                    \hline
                    Average Size of Group             &  2.08       &   7.19\\
                    \hline
                    Number of User-Item &  116344     &   3976\\
                    interactions & &\\
                    \hline
                    Number of Group-Item &  145068     &   3595\\
                    interactions & &\\
                    \hline
                    Interactive Repetition Rate       &  0.825      &   0.09\\
                    \hline
                \end{tabular}
                \end{minipage}
                \end{center}
            \label{Table 2}
            \end{table}
            
        \subsubsection{Evaluation Metrics}
            We adopted the "leave one" evaluation method, which has been widely used to evaluate the performance of Top@N recommendations. Specifically, for each group, we randomly delete several items in the interaction for testing, thereby dividing the training set and the test set, and the ratio of the training set to the test set is 10:1.
            
            For each group, we randomly select 100 items that have not been interacted with before as the negative sample. In order to evaluate the performance of Top@N recommendations, we have adopted widely used indicators Hit Rate (HR@N) and Normalized Discounted Cumulative Gain (NDCG@N).
            
            \begin{equation}
                HR@N=\frac{Hited}{N}\times 100\%
            \end{equation}
            \begin{equation}
             DCG@N=\sum_{n=1}^N\frac{pos_n}{log_2(n+1)}
            \end{equation}
            \begin{equation}
                NDCG@N=\frac{DCG@N}{IDCG@N}\times 100\%
            \end{equation}
            
             Where $pos_n$ represents the position of the item in the recommended list, and IDCG is an ideal situation for DCG.
            
             In the "Leave One" evaluation, HR measures whether the item used for testing is ranked in the "top@N" list (1 means yes, 0 means no), and NDCG is given by the ranking position of the item in the recommended list score. In order to facilitate comparison with existing work, we uniformly set the the experiment in HR@10, NDCG@10 and HR@5, NDCG@5.
        \subsubsection{Baselines}
            In order to show that our model is superior to the existing models(\hyperref[RQ 1]{RQ 1}), we compared the current superior models as follow:
            \begin{itemize}
                \item \textbf{NCF+AVG \cite{ref7,ref25}:} This is a model that aggregates scores. It uses the NCF framework for each group member to learn and predict the score of the item and regards the average score of the group member for the item as the group's score for the item. At the same time, there are also the maximum and minimum scores of group members as the group scores, but the effect is not as good as the average strategy, so only this method of aggregation is shown.
                \item \textbf{AGREE \cite{ref3}:} The model uses the attention mechanism, which determines the weight of each group member’s decision on the item according to the degree of interaction of each member in the group with the target item.
                \item \textbf{GREE:}This method is a variant of AGREE. It removes the attention mechanism of AGREE. It is assumed that each member has the same contribution to the group’s decision-making. It is different from NCF+AVG in that it calculates the group feature in advance and then calculates the score of the item for the group feature.
            \end{itemize}
            In order to illustrate the effect of group interaction repetition rate on recommendation performance(\hyperref[RQ 2]{RQ 2}), we set up the following experiment to verify our hypothesis in \hyperref[subsec4.3]{section 4.3}:
            \begin{itemize}
                \item \textbf{One Layer GNN:} Use only One Layer of the GNN mentioned in \hyperref[subsec4.3]{section 4.3} to observe the performance difference between the two datasets.
                \item \textbf{Two Layers GNN:} Use a two layesr GNN to compare the effect of using One Layer on two datasets.
                \item \textbf{Two Layers GNN with Residual Connection:} Compared with the ordinary Two Layers GNN, this residual layer is to directly add the output results of the two layers to verify whether the weighted two-layer network output will be better after the effect of the interactive repetition rate.
            \end{itemize}
        \subsubsection{Implementation and Hyperparameter Settings}
        
            We implement our method on the Pytorch platform and use the DGL library to implement our GNN model. We use the Adam \cite{ref8} optimizer to perform all gradient-based calculations. Based on experiments, we found that adjusting the learning rate to 0.05 can achieve the best results. Regarding the size of the data, we found that setting the Embedding size too large will increase the difficulty of training, that is, it is difficult to achieve convergence. If it is too small, it will lead to insufficient encoding of the necessary information. Therefore, we set the Embedding size to 32, and in two layers, the ReLU \cite{ref27} activation function is used between the GNN. In AGREE, GREE and NCF+AVG, the experimental settings we adopted refer to the best settings in \cite{ref3}. The hyperparameters in One/Two Layer GNN and Two Layers GNN with Residual Connection are consistent with our model parameters. We chose the xavier method for the initialization of the network, and the Embedding of each node uses the initialization based on the normal distribution. To prevent errors, we repeat each experiment 5 times, and the average value of the maximum value of each experiment plus the standard deviation is used as the final result of the model.
            
            \begin{figure}[ht]
                \centering
                \begin{minipage}[t]{\linewidth}
                    \centering
                    \includegraphics[scale=0.6]{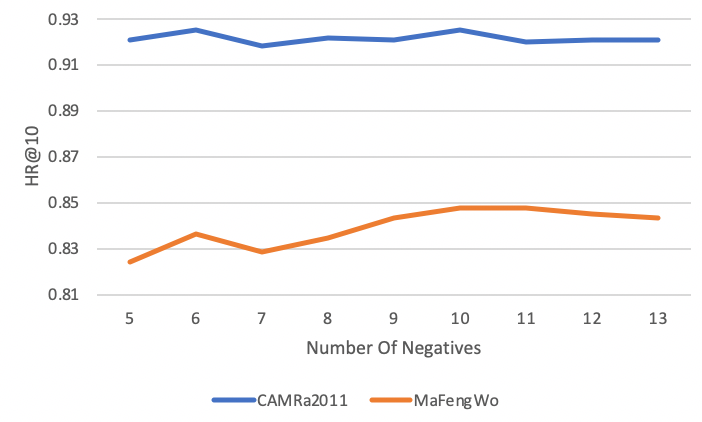}
                    \caption{Performance of HitRate@10 in the two datasets under different negative sampling ratios.}
                    \label{Figure 5}
                \end{minipage}
                \\
                \begin{minipage}[t]{\linewidth}
                    \centering
                    \includegraphics[scale=0.6]{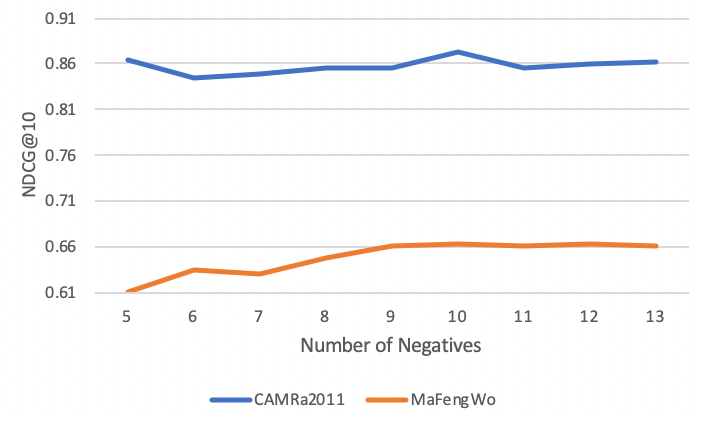}
                    \caption{Performance of NDCG@10 in the two datasets under different negative sampling ratios.}
                    \label{Figure 6}
                \end{minipage}
            \end{figure}
            
             Since the dataset does not have explicit negative feedback, 100 items that are not selected in each group are taken as negative samples. During training, we conducted a lot of experiments on these two datasets and found that the positive and negative sample ratio is adjusted between 1:10 and 1:12. The model works best. The specific adjustment process is shown in \hyperref[Figure 5]{Figure 5} and \hyperref[Figure 6]{Figure 6}. It can be seen that the changes in HR@10 and NDCG@10 in the CAMRa2011 dataset have not changed much since the beginning of 6, while the two indicators of the MaFengWo dataset have started to level off from 10. This is due to the problem of data scale. When the amount of data is relatively large and dense, a relatively low negative sampling rate will be required to make the model tend to fit. In summary, we set the positive-negative sample ratio to 1:10.
             
    \subsection{Performance Comparison}
        \subsubsection{Model Performance Comparison(\hyperref[RQ 1]{RQ 1})}
            \begin{table*}[htp]
            % \sidewaystablefn%
            \begin{minipage}{\textwidth}
            \centering
             \caption{\textbf{The performance of the deep neural network based model and our model on the CAMRa2011 dataset}}
                \begin{tabular}{@{\extracolsep{\fill}}ccccc@{\extracolsep{\fill}}}
                \hline
                     & \multicolumn{4}{c}{CAMRa2011} \\
                \hline
        & HR@10            & NDCG@10          & HR@5             & NCDCG@5 \\
                \hline
NCF+AVG & 0.753$\pm$ 0.072 & 0.443$\pm$ 0.052 & 0.554$\pm$ 0.032 & 0.376$\pm$ 0.044 \\
                
AGREE   & 0.787$\pm$ 0.035 & 0.456$\pm$ 0.042 & 0.572$\pm$ 0.028 & 0.384$\pm$ 0.038 \\
                
GREE    & 0.767$\pm$ 0.030 & 0.447$\pm$ 0.038 & 0.568$\pm$ 0.029 & 0.356$\pm$ 0.032 \\
                
GIP4GR    & \textbf{0.925$\pm$ 0.003} & \textbf{0.873$\pm$ 0.008} & \textbf{0.894$\pm$ 0.008} &           \textbf{0.861$\pm$ 0.020} \\
                \hline
                \end{tabular}
                \end{minipage}
            \label{Table 3}
        \end{table*}
        
        \begin{table*}[htp]
        \begin{minipage}{\textwidth}
            \centering
             \caption{\textbf{The performance of the deep neural network based model and our model on the MaFengWo dataset}}
                \begin{tabular}{@{\extracolsep{\fill}}ccccc@{\extracolsep{\fill}}}
                \hline
                     & \multicolumn{4}{c}{MaFengWo} \\
                \hline
        & HR@10            & NDCG@10          & HR@5             & NCDCG@5 \\
                \hline
NCF+AVG & 0.614$\pm$ 0.023 & 0.398$\pm$ 0.036 & 0.484$\pm$ 0.039 & 0.323$\pm$ 0.055 \\
                
AGREE   & 0.637$\pm$ 0.034 & 0.435$\pm$ 0.025 & 0.473$\pm$ 0.028 & 0.369$\pm$ 0.037 \\
                
GREE    & 0.615$\pm$ 0.031 & 0.414$\pm$ 0.024 & 0.442$\pm$ 0.032 & 0.348$\pm$ 0.031 \\
                
GIP4GR    & \textbf{0.848$\pm$ 0.010} & \textbf{0.662$\pm$ 0.004} & \textbf{0.750$\pm$ 0.013} &           \textbf{0.609$\pm$ 0.023} \\
                \hline
                \end{tabular}
            \end{minipage}
            \label{Table 4}
            \end{table*}

        \hyperref[Table 3]{Table 3} and \hyperref[Table 4]{Table 4} represent the performance on the datasets CAMRa2011 and MaFengWo, respectively. It can be seen that our proposed model performs much better on HitRate and NDCG than the existing models. Among them, NCF+AVG has the worst effect. (NCF+AVG, AGREE, GREE three models based on ordinary neural networks have not very different results in these indicators). Our model uses GNN, each embedding of a node gathers more information from surrounding nodes and can learn more effectively than ordinary neural networks. Therefore, the results of training the model also show a lower variance, indicating that the performance of the model is also relatively stable. At the same time, the method based on the characteristics of group members (AGREE, GREE) is better than relying solely on scores (NCF+AVG). This shows that the degree of preference for an item of each group member cannot reflect the preferences of items when gathered in a group. Finally, the model with the Attention mechanism (AGREE) is better than the average of group member characteristics (GREE), which shows that different group members have different contributions to group decision-making, so it also explains to us the graph's attention network (GAT) is used when fusing individuals into group features.
        
        \subsubsection{The Effect of IRR on Model Performance(\hyperref[RQ 2]{RQ 2})}
        
            \begin{table*}[htp]
            \begin{minipage}{\textwidth}
            \centering
             \caption{\textbf{Comparison of the performance of the general GNN model and our model on the CAMRa2011 dataset}}
                \begin{tabular}{@{\extracolsep{\fill}}ccccc@{\extracolsep{\fill}}}
                \hline
                     & \multicolumn{4}{c}{CAMRa2011} \\
                \hline
            & HR@10            & NDCG@10          & HR@5             & NCDCG@5 \\
                \hline
One Layer GNN     & 0.920$\pm$ 0.004 & 0.860$\pm$ 0.012 & 0.892$\pm$ 0.010 & 0.837$\pm$ 0.001 \\
                
Two Layers GNN    & 0.867$\pm$ 0.017 & 0.612$\pm$ 0.055 & 0.739$\pm$ 0.033 & 0.582$\pm$ 0.050 \\
                
One Layer GNN+Res & 0.875$\pm$ 0.016 & 0.613$\pm$ 0.042 & 0.756$\pm$ 0.033 & 0.582$\pm$ 0.050 \\
                
GIP4GR    & \textbf{0.925$\pm$ 0.003} & \textbf{0.873$\pm$ 0.008} & \textbf{0.894$\pm$ 0.008} &           \textbf{0.861$\pm$ 0.020} \\
                \hline
                \end{tabular}
                \end{minipage}
            \label{Table 5}
            \end{table*}
        
            \begin{table*}[htp]
            \begin{minipage}{\textwidth}
            \centering
             \caption{\textbf{Comparison of the performance of the general GNN model and our model on the MafengWo dataset}}
                \begin{tabular}{@{\extracolsep{\fill}}ccccc@{\extracolsep{\fill}}}
                \hline
                     & \multicolumn{4}{c}{MaFengWo} \\
                \hline
            & HR@10            & NDCG@10          & HR@5             & NCDCG@5 \\
                \hline
One Layer GNN     & 0.741$\pm$ 0.012 & 0.543$\pm$ 0.011 & 0.631$\pm$ 0.018 & 0.508$\pm$ 0.018 \\
                
Two Layers GNN    & 0.836$\pm$ 0.018 & 0.638$\pm$ 0.032 & 0.732$\pm$ 0.027 & 0.594$\pm$ 0.037 \\
                
One Layer GNN+Res & 0.838$\pm$ 0.011 & 0.643$\pm$ 0.017 & 0.735$\pm$ 0.018 & 0.581$\pm$ 0.019 \\
                
GIP4GR    & \textbf{0.848$\pm$ 0.011} & \textbf{0.662$\pm$ 0.004} & \textbf{0.750$\pm$ 0.013} &           \textbf{0.609$\pm$ 0.023} \\
                \hline
                \end{tabular}
                \end{minipage}
            \label{Table 6}
            \end{table*}

        It can be seen from \hyperref[Table 5]{Table 5} and \hyperref[Table 6]{Table 6} that due to the relatively large IRR on the CAMRa2011 dataset, the One Layer GNN has better performance than the Two Layers GNN; and the MaFengWo dataset is due to the comparison of the IRR is small, so using a Two Layers GNN is better than using a One Layer GNN.
        But we pursue a more universal model structure, it is impossible to use different models for different datasets. Secondly, the effect of Two Layers+Res is more like a compromise between the simple use of One Layer and Two Layers GNN. It simply adds the output of two layers. The price of its universality is its performance. This approach is similar to JKnet, that is, let the model learn the weights of different layers of each node. Due to the sparseness of the data, the performance of the model cannot be optimal and the performance is unstable (that is, the standard deviation is relatively large.) so it is difficult to converge during training, and our proposed model has achieved the best results by fusing the features between layers according to the structure of the dataset.
        
        It is worth noting that comparing \hyperref[Table 3]{Table 3} and \hyperref[Table 5]{Table 5}, \hyperref[Table 4]{Table 4} and \hyperref[Table 6]{Table 6} respectively, it can be found that the model performance of GNN based on single layer or double layer is better than the model based on deep neural network, which also shows in the group-user-item interaction graph, there is more structural information that can be obtained through GNN, so the use of GNN in group recommendation is the correct choice.
        
        \begin{figure}[ht]
                \centering
                \includegraphics[scale=0.3]{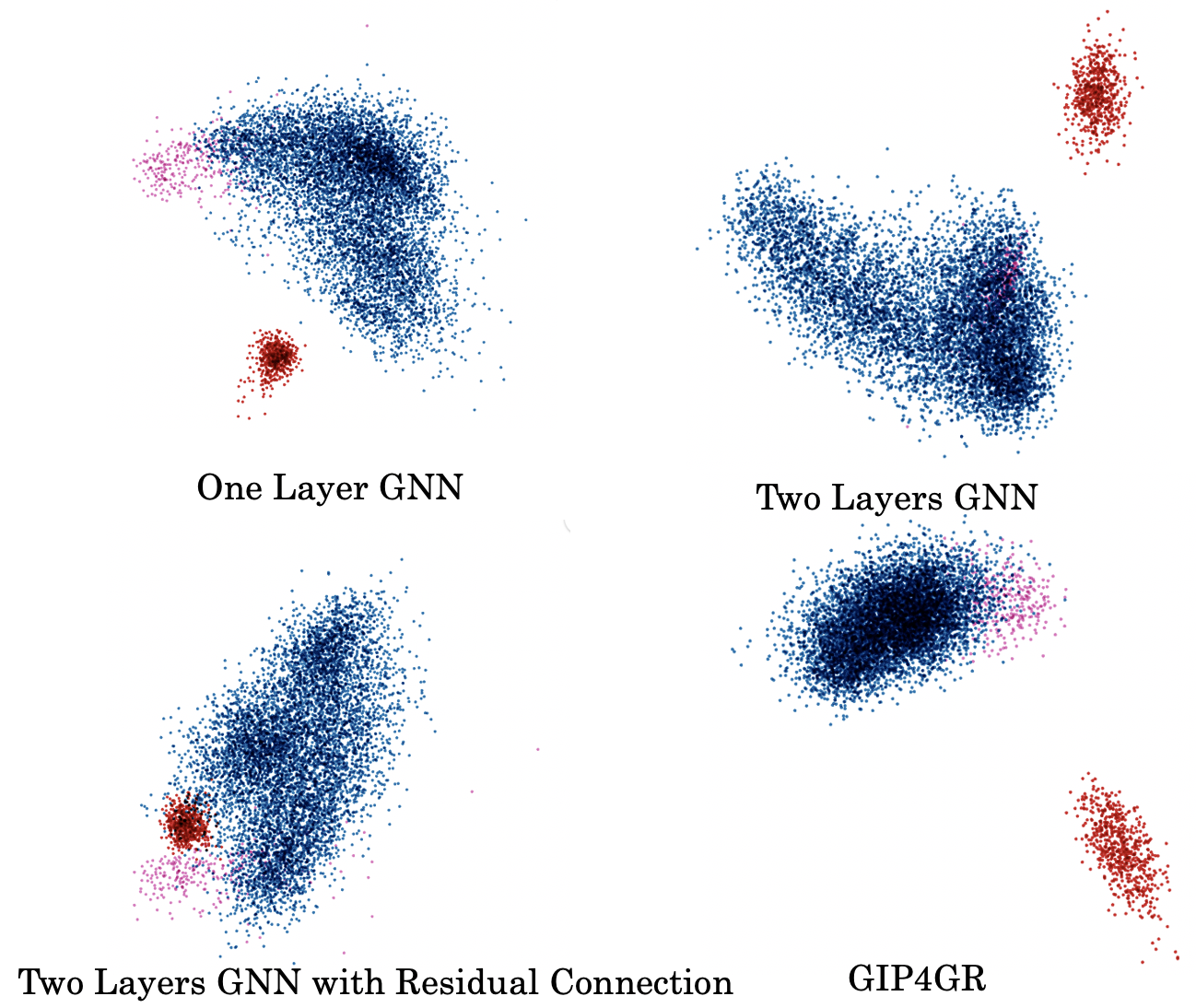}
                \caption{CAMRa2011 dataset's dimensionality reduction visualization.}
                \label{Figure 7}
            \end{figure}
            
            Finally, the representation of the trained nodes embedding is reduced by T-SNE, and conclusions similar to the above results can be obtained intuitively. In \hyperref[Figure 7]{Figure 7}, the blue dots represent items, the red dots represent users, and the pink dots represent groups. It can be seen from \hyperref[Figure 7]{Figure 7} that due to the impact of the IRR of the dataset, the effect of One Layer of GNN and our model GIP4GR are relatively close, and the Two Layers GNN with Residual Connection model is inferior to them, but the three points are separate, you can't make the difference significant. The Two Layers GNN have the worst effect, which is reflected in the fact that the embedding of the group and the embedding of the item are difficult to separate, which is consistent with our experimental data.
            
            \begin{figure}[H]
                \centering
                \includegraphics[scale=0.3]{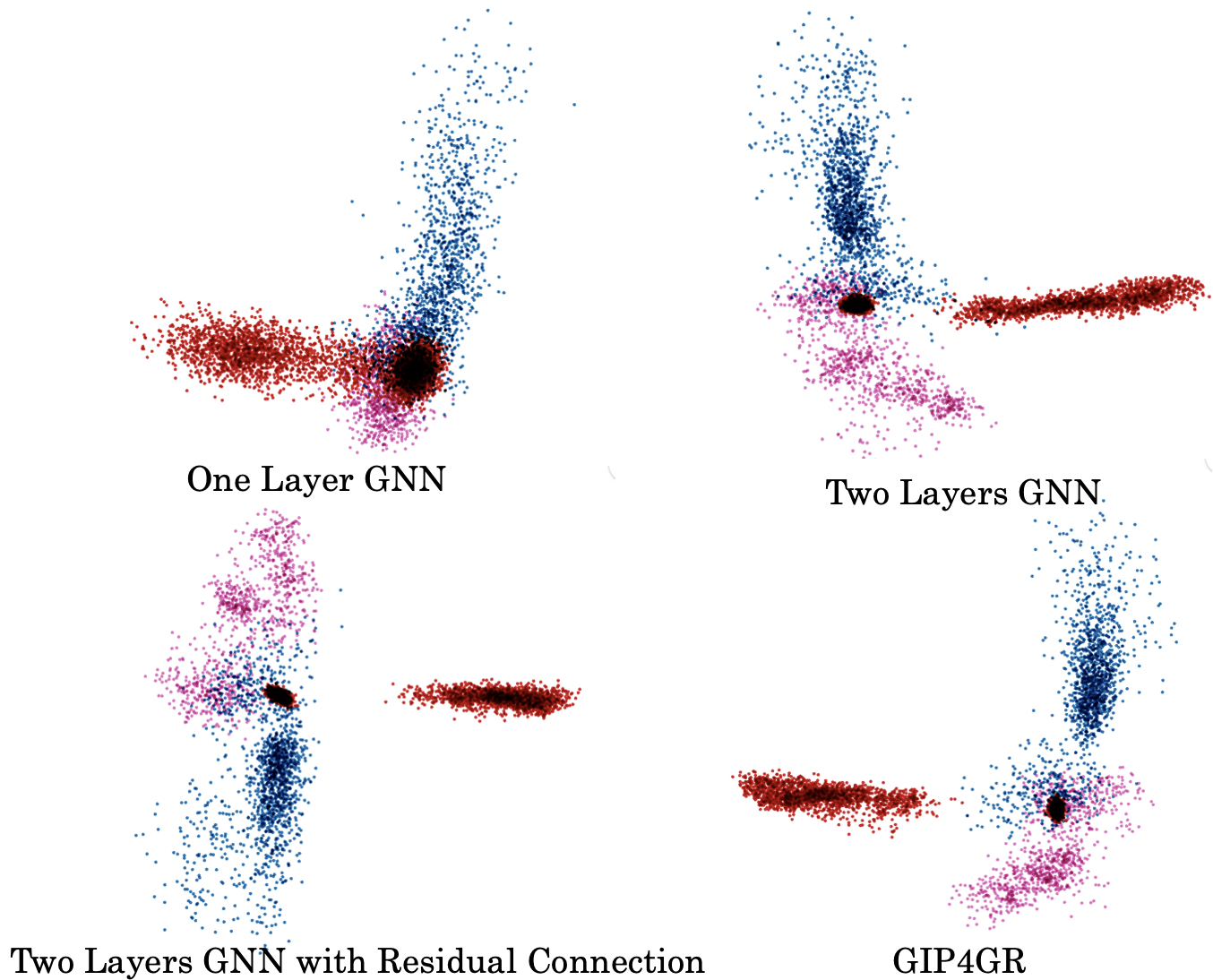}
                \caption{MaFengWo dataset's dimensionality reduction visualization.}
                \label{Figure 8}
            \end{figure}
            
            Meanwhile, in the MaFengWo dataset, it can be seen intuitively from \hyperref[Figure 8]{Figure 8} that due to the relatively small IRR, the effect of using a One Layer GNN is the worst. Secondly, the effects of the Two layers GNN and our model GIP4GR are relatively close, and the effect of the Two Layers GNN with Residual Connection is inferior to the two, which is consistent with our hypothesis.

\section{Conclusion}\label{sec6}
In this work, we proposed a GNN method to solve the problem of group recommendation and proposed a new model GIP4GR. First, we put forward the concept of group interaction repetition rate for the problem of group recommendation. Secondly, we proposed the use of GNN to learn on the group recommendation dataset, that is, on each layer of the network, GAT is used to learn the contribution of each member in the group decision-making, and GraphSAGE is used to aggregate the group and the item, the user's representation. Finally, the group IRR is applied to a GNN-based framework, and the characteristics of the dataset itself are used to weighting the first layer and second layer GNN for better performance. And a large number of experiments have been conducted on two real-world datasets and compared with the best existing methods, our model has achieved the best results. The current shortcoming of this work is that it does not consider the interaction between members of the group, that is, the members are only regarded as the attribute nodes of the group, and the interactive connections of the group members are not counted as part of the graph. Subsequent work may consider connecting the member nodes in the group, separately considering the message passing between users when the entire interaction graph is used for message passing, and reflecting this influence on the weight of the group are two important issues.

\section{Declarations}
\textbf{Conflict of interest} The authors declare that they have no conflict of interest.


\bibliography{sn-bibliography}% common bib file


\end{document}